\documentclass[prb,twocolumn,longbibliography,amsmath,superscriptaddress,amssymb]{revtex4-2}

\usepackage{graphicx}
\usepackage{dcolumn}
\usepackage{bm}
\usepackage{amssymb}
\usepackage{amsmath}
\usepackage{color}
\usepackage[dvipsnames]{xcolor}
\usepackage{placeins}
\usepackage{epstopdf}

\usepackage[normalem]{ulem}
\usepackage[linkcolor=blue,urlcolor=blue,citecolor=blue,colorlinks=true]{hyperref}

\newcommand{\bra}[1]{\langle\,#1\,|}
\newcommand{\ket}[1]{|\,#1\,\rangle}
\newcommand{\braket}[2]{\langle\,#1\,|\,#2\,\rangle}
\newcommand{\bss} {Bi$_2$Se$_3$}
\newcommand{\btt} {Bi$_2$Te$_3$}
\newcommand{\eq} [1]{Eq.~(\ref{#1})}

\def\ev{\boldsymbol\epsilon} 
\def\uu{\uparrow}
\def\dd{\downarrow}
\def\yy{\mathbf y}                                       
\def\xx{\mathbf x}                                       
\def\nn{\mathbf{n}}                                       
\def\sx{\mathbf{s}}                                       
\def\kk{\mathbf k}
\def\rr{\mathbf{r}}                                       

\def\RR{\mathbf{r}_\parallel}
\def\KK{\mathbf{k}_\parallel}
\def\KA{k_\parallel}
\def\GV{\mathbf{G}_\parallel}
\def\TG{\boldsymbol\tau}
\def\KG{\boldsymbol\kappa}
\def\XG{\boldsymbol\xi}
\def\EG{\boldsymbol\eta}
\def\KP{${\mathbf k}\!\cdot\!{\mathbf p}$}

\begin{document}
\title{Theory of circular dichroism in angle- and spin-resolved photoemission from the surface state on Bi(111)}

\author{E. E. Krasovskii}
\affiliation{Donostia International Physics Center (DIPC), 20018 Donostia/San Sebasti\'{a}n, Basque Country, Spain}
\affiliation{Departamento de Pol\'imeros y Materiales Avanzados: F\'isica, Qu\'imica y Tecnolog\'ia, Universidad del Pais Vasco/Euskal Herriko
Unibertsitatea, 20080 Donostia/San Sebasti\'an, Basque Country, Spain}
\affiliation{IKERBASQUE, Basque Foundation for Science, 48013 Bilbao, Basque Country, Spain}

\begin{abstract}
  Circular dichroism (CD) of photoemission from a surface state on Bi(111) is considered
  within an {\it ab initio} one-step theory of photoemission. The symmetry of the azimuthal
  distribution of CD for different spin directions is discussed in terms of the spin structure
  of the initial states. Both normal and off-normal light incidence are considered. Strong
  photon energy dependence of photoemission in the ultraviolet range is observed. The principal
  difference between the initial state spin polarization and the spin photocurrent is emphasized
  and problems with revealing the orbital angular moment with photoemission are pointed to.
\end{abstract}

\maketitle

\section{Introduction}
Photoemission has been long established as the most straightforward method to infer the
electronic structure of a crystal~\cite{Huefner,Schattke-van_Hove}. The energy and angular
distribution of photoelectrons brings information about the surface-projected spectral
function. A further degree of freedom is the spin of the photoelectron. For nonmagnetic
solids a spin dependence of the photocurrent in vacuum characterizes relativistic
effects~\cite{Dil2009,Meier_2009,Heinzmann2012,Okuda2013,Krasovskii2015,Okuda2017}.
The theoretical background is given by the one-step theory of photoemission
\cite{Adawi64,Mahan1970,Caroli73,Feibelman1974,Pendry1976}, which was applied to various
aspects of spin-resolved photoemission~\cite{BORSTEL1985,Feder1985,Braun_1996,Braun2018};
in particular to the symmetry analysis of the spin photocurrent~\cite{Tamura1987,HENK1998}.
For a specific
initial state, the photocurrent depends on the geometric parameters of the photoemission
setup---light polarization and incidence geometry, as well as the photon energy $\hbar\omega$.
By varying these parameters one can reveal certain aspects of the structure of the initial
state wave function.

In particular, the idea was put forward that measurements with circularly polarized light may
provide information about orbital angular moment
(OAM)~\cite{Jung2011OAM,Park2012OAM,Kim2012OAM,Cho2018,Schueler2020}.
The photoemission CD was claimed to be related to OAM for the surface states in \btt
\cite{Jung2011OAM}, \bss \cite{Park2012OAM,Bahramy2012}, Cu(111) and Au(111)~\cite{Kim2012OAM}.
This interpretation was questioned by Scholz~{\it et al.}~\cite{Scholz2013} based on the
apparent photon energy dependence of CD---in particular, a reversal of the sign was observed
in \btt. Indeed, $\hbar\omega$-dependence of CD has been widely observed experimentally for
various materials: Bi$_2$Te$_2$Se~\cite{Neupane2013}, Bi$_2$Se$_3$~\cite{Barriga2014},
BiTe$X$, $X=\rm I$, Br, Cl~\cite{Crepaldi2014}. Furthermore, a theoretical analysis of
CD of the surface states on Au(111) with the one-step photoemission theory~\cite{Arrala2013}
showed rapid changes of CD with $\hbar\omega$.

Regarding the spin photocurrent, some authors expected it to follow the spin polarization of
the initial states~\cite{Takayama2011,Pan2011,Hoepfner2012}. At the same time, the spin
photocurrent is known to depend on the light polarization and experimental
geometry~\cite{ParkCH2012,Jozwiak2013,Zhu2013,Henk2003,Bentmann2017}.
Ref.~\cite{Mirhosseini2012} admits the important role of final-state effects but concludes
that CD is a suitable method to analyze spin texture. In a simple pseudospin
model~\cite{Wang2011}, CD can be related to the spin polarization of the initial states,
and the azimuthal dependence of circular dichroism from the topological surface states (TSS)
in \bss\ was experimentally observed in refs.~\cite{Park2012OAM} and~\cite{Wang2011} to
follow the spin texture of the TSS model. On the contrary, the theoretical analysis in
Ref.~\cite{Scholz2013} established a final-state origin of CD in \btt\ and demonstrated that
the spin photocurrent was not related to the sign of CD. This point of view was confirmed
experimentally for \bss\ in Ref.~\cite{Barriga2014}. There, over a wide $\hbar\omega$ interval
the spin texture was found to be independent of light polarization, from which the authors
concluded that the photoelectron spin polarization was ``representative of the initial TSSs''.

The present paper theoretically addresses the above issues for the surface state on the Bi(111)
surface with the aim to reveal possible difficulties that may prevent CD from being interpreted
solely in terms of properties of initial states. Here, the azimuthal distribution of
spin-resolved CD and photocurrent spin polarization in the photon-energy interval
$\hbar\omega=7-19$~eV is studied using an {\it ab initio} one-step theory of photoemission.
In this proof-of-principle calculation we are interested in the symmetry of the spin-dependent
CD distribution and its $\hbar\omega$ dependence. The Bi(111) surface is interesting because
of the very strong spin-orbit coupling (SOC), which makes it of practical importance for
spin-to-charge current conversion~\cite{Isasa2016}. It is particularly instructive owing
to its $C_{3v}$ symmetry, which is common to many topological insulators, transition-metal
dichalcogenides, as well as elementary metals---popular objects of spectroscopic studies.

\vfill

\begin{figure}[b] 
\centering
\includegraphics[width=0.99\columnwidth]{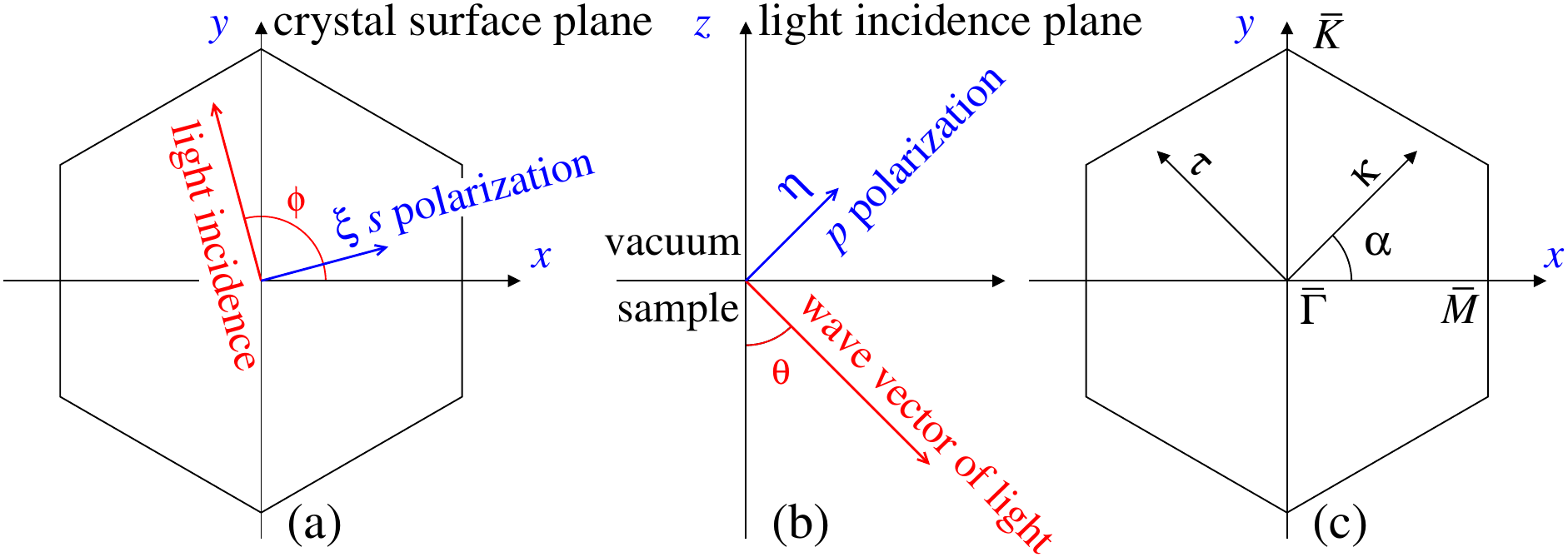} 
\caption{Light incidence geometry and local coordinate frame. (a)~Crystal surface projection
  of the light incidence plane at the azimuth $\phi$ and $s$-polarization vector $\XG$.
  (b)~Light incidence angle $\theta$ and $p$-polarization vector $\EG$. (c)~Local coordinate
  frame at the $\KK$ azimuth $\alpha$.
}
\label{f:light}
\end{figure}
\section{Theoretical background and computational methodology}
Angle-resolved photocurrent is calculated based on the one-step theory of photoemission
\cite{Adawi64,Mahan1970,Caroli73,Feibelman1974,Pendry1976}, in which
the optical excitation of the electron, its scattering by the
crystal-vacuum interface, and transport to the detector are subsumed
into the time reversed LEED (low energy electron diffraction)
state. The photocurrent in vacuum excited by light of frequency
$\omega$ from the initial state $\ket{\rm{i}\,\KK}$ reads
\begin{equation}
I(\omega,\KK)\sim \sqrt{E_{\rm f}-E_{\rm vac}}\,
\left|\bra{{\rm f}\,\KK}\hat o\ket{\rm{i}\,\KK}\right|^2.
\label{photocurrent}
\end{equation}
Here $\braket{\rr}{\rm{f}\,\KK}\equiv\Phi(\rr)$ is the time reversed LEED wave
function at energy $E_{\rm f}=E_{\rm i}+\hbar\omega$ and surface-parallel crystal momentum
$\KK$. $E_{\rm vac}$ is the vacuum level (kinetic energy zero). Equation~(\ref{photocurrent})
stems from the well-established time-dependent perturbation theory, so the accuracy of the
photocurrent depends on the accuracy of the wave functions of the quasi-particle states
$\ket{{\rm f}\,\KK}$ and $\ket{\rm{i}\,\KK}$.
Here, both the initial and the final states are eigenfunctions of the Kohn-Sham Hamiltonian
obtained within the local density approximation of the density functional theory.

The interaction with the photon field is calculated in the non-relativistic dipole
approximation, i.e., the spin-orbit related term~\cite{Feder1985} is neglected. The validity
of this approximation is discussed in Ref.~\cite{Krasovskii2015}. Thus, the perturbation reads
$\hat o ={\boldsymbol -i\nabla\cdot\ev}$, where $\ev$ is a complex polarization vector.
Circular dichroism is the difference between the intensities for circular right (CR) and
circular left (CL) light polarizations, with $\ev_{\rm CR}=\XG-i\EG$ and $\ev_{\rm CL}=\XG+i\EG$, 
where $\XG$ and $\EG$ are two orthogonal unit vectors in the plane perpendicular to the light
propagation, see Figs.~\ref{f:light}(a) and~\ref{f:light}(b), and it arises from the
interference of two transition amplitudes: $\ev = \XG$ and $\ev=\EG$. Note that here the
dichroism is defined as the difference $D=I_{\rm CR}-I_{\rm CL}$, which is not normalized to
the total intensity $I=I_{\rm CR}+I_{\rm CL}$. This is convenient when one considers the $\KK$
distribution of the dichroism because it gives a better idea of what would be the result of
the angular averaging.

The LEED wave function describes the scattering of the plane wave $\exp(-i\kk\rr)$ incident
from vacuum on the surface, where $\kk=\KK+\nn\sqrt{2mE/\hbar^2-\KA^2}$. In the semi-infinite
crystal $\Phi$ satisfies the Schr\"odinger equation $\hat H\Phi=E\Phi$, with the
scalar-relativistic Hamiltonian $\hat H$, which contains the imaginary potential $-iV_{\rm i}$
to allow for the inelastic scattering of the outgoing electron. Here, a constant value
$V_{\rm i}=1$~eV is assumed. 
\begin{figure}[t] 
\centering
\includegraphics[width=0.95\columnwidth]{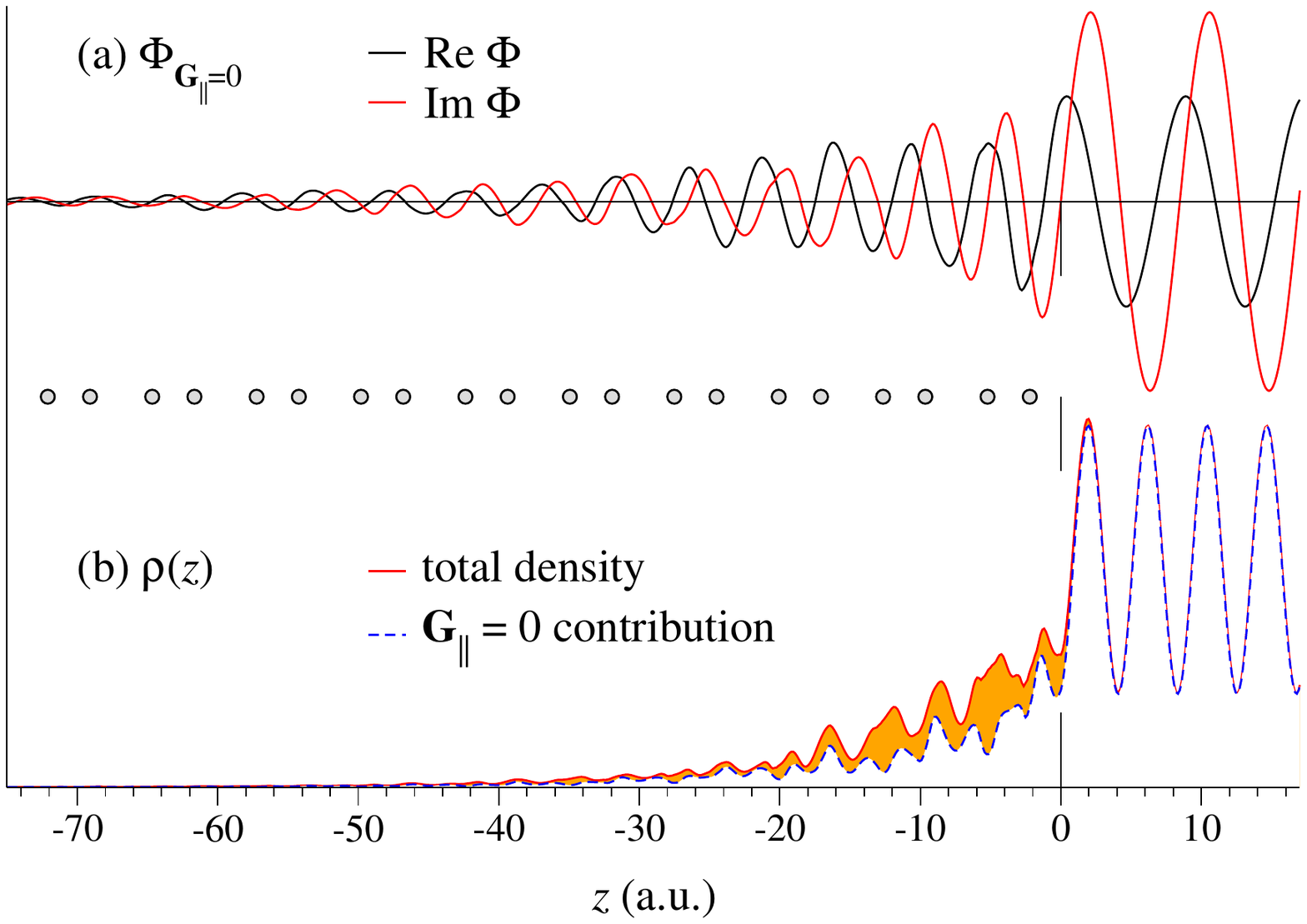} 
\caption{(a)~$\GV=0$ Fourier component of the time-reversed LEED function for $\KK=0$ and
$E_{\rm f}=10$~eV. (b) Probability density profile $\rho(z)$: red line is total density, see
Eq.~(\ref{profile}), and dashed blue line is the $\GV=0$ contribution $|\phi_{\GV=0}(z)|^2$.
Orange shaded area shows the contribution from the $\GV\ne0$ surface Fourier harmonics.
Circles show the location of the Bi(111) atomic planes. The matching plane between the
semi-infinite crystal and vacuum is at $z=0$.}
\label{scheme}
\end{figure}

In the LEED calculation the electron is incident from the right, and in the crystal half-space
$z<0$ the wave function $\Phi(\rr)$ is represented by a sum of the Bloch eigenstates,
both propagating and evanescent. The set of the Bloch solutions for a given $E_{\rm f}$ and
$\KK$ is obtained with the inverse \KP\ method in terms of augmented plane
waves~\cite{Krasovskii1997}. At the matching plane $z=0$, see Fig.~\ref{scheme}, the sum of
the bulk solutions is matched in value and slope to a sum of the incident and reflected plane
waves in vacuum, as explained in Ref.~\cite{Krasovskii1999}. Thus, the time-reversed LEED
wave function fully takes into account the singularity of the crystal potential at the nuclei
and its variation in the interstitial. Finally, the Laue representation of the final state is
obtained by a straightforward expansion of the wave functions in terms of plane waves:
\begin{equation}
\Phi(\RR,z) = \sum\limits_{\GV} \phi_{\GV}(z)\exp[i(\KK+\GV)\,\RR],
\label{laue}
\end{equation}
where $\RR$ is the position vector parallel to the surface and $\GV$ are surface reciprocal
lattice vectors. An example of the scattering solution for the normal emission at $E=10$~eV
relative to the Fermi energy is presented in Fig.~\ref{scheme} as the zeroth Fourier component
$\phi_{0}(z)$ and density profile

\begin{equation}
\rho(z) = \int|\,\Phi(\RR,z)\,|^2\, d\RR.
\label{profile}
\end{equation}

Note that in the interior of the crystal the contribution from the Fourier harmonics with
reciprocal lattice vectors $\GV\ne0$ (orange) is comparable to the $\GV=0$ contribution,
which points to possible problems with applying a free-electron approximation for the
final state~\cite{Krasovskii2020}. 

\section{Spin structure of the Bi(111) surface state}\label{s:ground}
\begin{figure}[b!] 
\centering
\includegraphics[width=0.99\columnwidth]{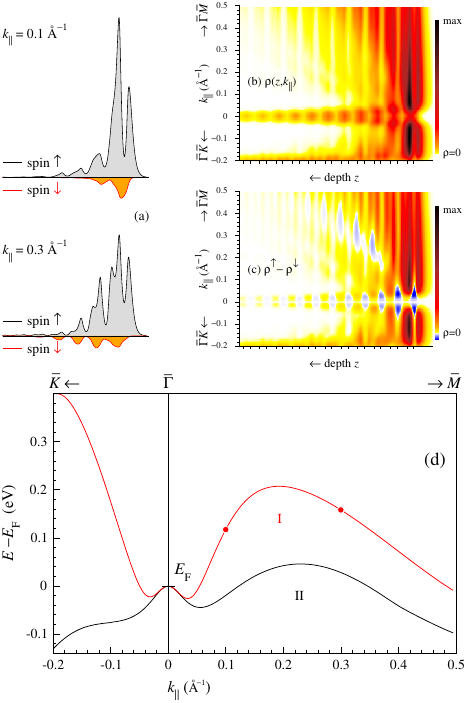} 
\caption{Density profiles and $\KK$ dispersion of the surface states on Bi(111).
  (a)~Spin-density profiles of the surface state~I in the $\bar\Gamma\bar M$ line, at
  $\KA=0.1$ and 0.3~\AA$^{-1}$ [circles in graph (d)]. Spin axis $\sx$ is along $\yy$,
  see Fig.~\ref{f:light}(c). (b)~Total density distribution for this state for $\KK$
  along $\bar\Gamma\bar M$ and $\bar\Gamma\bar K$.
  (c)~Net-spin density distribution for $\sx\perp\KK$. Blue color
  means that minority spin density is larger than majority at this $z$.
The ticks in the $z$ axis of graphs (b) and (c) denote the atomic planes. 
  (d)~Energy dispersion $E(\KK)$ of the two surface states.
}
\label{f:sus}
\end{figure}

Spin-orbit effects at surfaces are often discussed in terms of the spin direction of the
electronic states. In the spirit of the Rashba model, much attention is paid to the details of
the spin-momentum locking and its relation to spin-polarized photocurrent~\cite{Usachov2020}.
However, in real crystals the wave functions are generally not spin eigenfunctions, i.e., the
spin direction varies from one real space point to another. A direct consequence of this is the
dependence of the spin photocurrent on the geometrical setup of the photoemission measurement
and on the photon energy~\cite{Arrala2013,Scholz2013,Bentmann2017}, as we will see in the next
section. Both the spin polarization of the photocurrent and dichroism depend on the spinor
structure of the initial state owing to the different dependence on $\ev$ of the matrix
elements $\bra{{\rm f}\,\KK}\hat o \ket{{\rm i}^\uu\,\KK}$ and
$\bra{{\rm f}\,\KK}\hat o \ket{{\rm i}^\dd\,\KK}$. It is most convenient to consider the
properties of the spinor components in the local coordinate frame, see Fig.~\ref{f:light}(c):
for a given $\KK$ azimuth we introduce unit vectors $\KG$ along $\KK$ and $\TG=\nn\times\KG$,
where $\nn$ is the surface normal. Then, for the spin quantization direction $\sx\parallel\TG$,
as zeroth approximation, one may start with the solution for a semi-infinite jellium (Rashba
model) and consider the spin-$\uu$ component to be an even function under the transformation
$\tau\to-\tau$ and the spin-$\dd$ one to be negligible. Figure~\ref{f:sus} demonstrates that
the spin structure of the Bi(111) surface state substantially deviates from this simple picture.

The surface states are calculated as eigenfunctions of a repeated-slab crystal, with a
70~\AA~wide supercell that comprises a 32-layer Bi(111) slab and a 9~\AA~vacuum clearance.
The crystal potential is artificially perturbed at one of the surfaces in order to separate
the surface states at the opposite surfaces and create physically relevant non-degenerate
states localized at the other surface, see Fig.~\ref{f:sus}(a). The calculations are performed
with the self-consistent full-potential linear augmented plane waves method~\cite{KSS1999}.
Spin-orbit coupling is included within the second-variational two-component
Koelling-Harmon approximation~\cite{KOH77,MCD80}, with the potential gradient being
taken into account only in the muffin-tin spheres.

For a given $\KK$, the slab eigenstates are a set of discrete levels, which are related to
bulk or to surface states of the true semi-infinite crystal depending on how strongly they are
localized. Figure~\ref{f:sus}(d) shows the dispersion $E(\KK)$ of the two levels that represent 
the two spin-orbit-split surface states. The $E(\KK)$ curves along $\bar\Gamma\bar M$ and
$\bar\Gamma\bar K$ agree well with the seminal calculation of Ref.~\cite{Koroteev2004}. In the
vicinity of the $\bar\Gamma$ point the surface states merge with the bulk continuum, which in
the slab calculation is reflected in the extended character of the density profile $\rho(z)$
in Fig.~\ref{f:sus}(b) at small $\KA$. In the $\bar\Gamma\bar K$ direction the upper state
disperses steeply to merge with bulk continuum at around $\KA=0.2$~\AA$^{-1}$, while along
$\bar\Gamma\bar M$ it remains localized over much wider $\KA$ range.

\begin{figure*}[t] 
\centering
\includegraphics[width=0.95\textwidth]{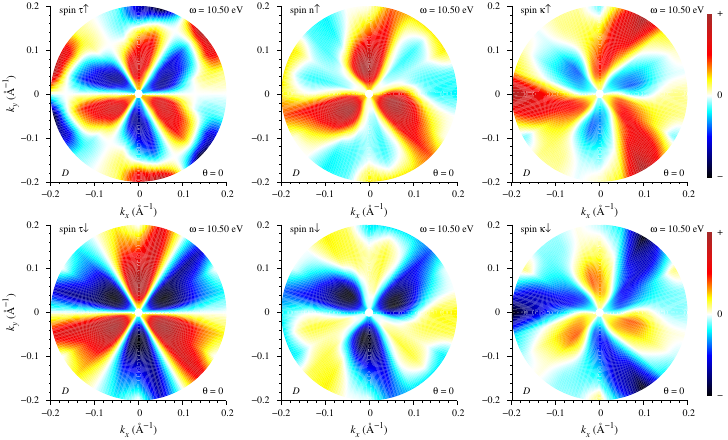} 
\caption{Normal incidence CD for $\hbar\omega=10.5$~eV for three spin directions in the local
  $\KK$-dependent frame, see Fig.~\ref{f:light}(c): $\sx\parallel\TG$ (left column),
  $\sx\parallel\nn$ (central column), and $\sx\parallel\KG$ (right column). Upper row shows
  spin-$\uu$ and lower row spin-$\dd$ CD. Maps are normalized to the maximum magnitude.
}
\label{f:ni1}
\end{figure*}
Figure~\ref{f:sus}(a) shows the spin density profiles for two states in the $\bar\Gamma\bar M$
line. It lies in the mirror plane $xz$ of the (111) surface, see Fig.~\ref{f:light}(c), and for
the spin axis along $\yy$ the two spinor components are eigenfunctions of different parity of
the reflection operator $y\to-y$~\cite{Henk2003}. For the higher energy state in
Fig.~\ref{f:sus}(d) the majority spin component $\psi^\uu$ is even and the minority one
$\psi^\dd$ is odd; the spin density profiles are shown in Fig.~\ref{f:sus}(a). Minority spin
comprises some 10 to 17\% of the total charge, and the spin polarization
\begin{equation}
S = \int |\,\psi^{\uu}(\rr)\,|^2-|\,\psi^{\dd}(\rr)\,|^2\, d\rr
\label{e:pol}
\end{equation}
does not exceed 80\%. Similar values occur for the $\bar\Gamma\bar K$ direction, i.e., for
$\KK$ along $\yy$ and spin axis along $-\xx$. At the same time, locally the spin minority
density may exceed the majority one, as illustrated in the color map Fig.~\ref{f:sus}(c),
e.g., around $\KA=0.3$~\AA$^{-1}$ along $\bar\Gamma\bar M$.
\begin{figure*}[ht] 
\centering
\includegraphics[width=0.95\textwidth]{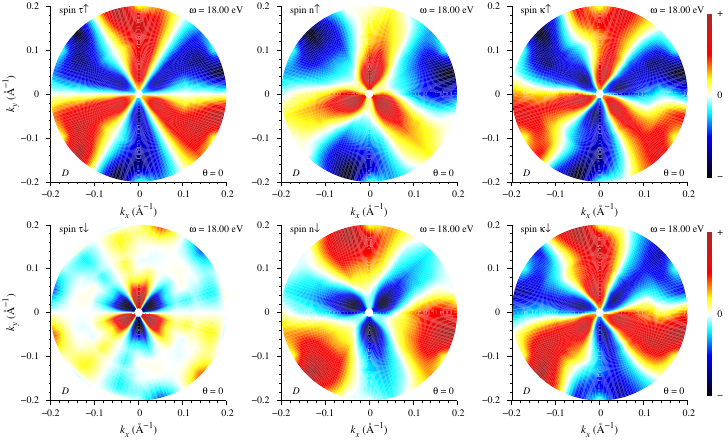} 
\caption{The same as in Fig.~\ref{f:ni1} for $\hbar\omega=18$~eV.}
\label{f:ni2}
\end{figure*}

\section{Results and discussion}
Let us focus on the light-polarization dependence of the emission intensity from the upper
surface state. In spite of the considerable deviation from the spin structure in the Rashba
model, for the spin axis perpendicular to $\KK$ the spin-up component is significantly larger
than spin-down. In the next sections we will consider the circular
dichroism for the three directions of the spin axis: $\TG$, $\KG$, and $\nn$.

\subsection{Normal light incidence}
The $\KK$ dependence of the dichroism of photoemission at normal incidence for three spin
directions is shown in Figs.~\ref{f:ni1} and~\ref{f:ni2} for $\hbar\omega=10.5$ and 18~eV,
respectively. The color maps are seen to have different symmetry depending on the spin
quantization axis $\sx$: for $\sx\!\parallel\!\TG$ the maps are antisymmetric on reflection in
the $\bar\Gamma\bar M$ plane: $D^{\uu\dd}_{\TG}(\alpha)=-D^{\uu\dd}_{\TG}(-\alpha)$ (here subscript
$\TG$ indicates the choice of the spin quantization axis), and the maps for opposite spins are
independent. For $\sx\!\parallel\!\KG$ and $\sx\!\parallel\!\nn$ the maps do not possess the
mirror symmetry, but the dichroic signals for the opposite spin directions are related:
$D^{\uu}_{\sx}(\alpha)=-D^{\dd}_{\sx}(-\alpha)$.

The total photocurrent CD, i.e., the sum of the maps $D=D^{\uu}_{\sx}+D^{\dd}_{\sx}$ is
independent of the $\sx$ direction, and it has the same symmetry as the $D^{\uu\dd}_{\TG}$ maps.
For $\KK$ sufficiently close to $\bar\Gamma$ these maps are approximately antisymmetric across
the $\bar\Gamma$ point (resembling the symmetry of the ground-state spin texture), but at larger
$\KA$ the deviation from the relation $D(\alpha+\pi)\approx-D(\alpha)$ becomes ever stronger. 

For all the choices of the spin axis CD is seen to strongly depend on the magnitude
of the $\KK$ vector: it may change sign, and this happens at different $\KA$ for different
$\alpha$. Generally, the structure of all the maps is rather complicated, which reflects the
complicated spatial and spin structure of the surface state. Furthermore, the maps strongly
depend on the photon energy $\hbar\omega$. Nevertheless, gross symmetry features of the maps
can be understood from a simplistic model that neglects the $z$-coordinate of the wave
functions. Then the general expression for the initial state spinor along the spin-axis $\tau$
of the local coordinate frame $(\kappa,\tau)$ reads

\begin{equation}
\psi_\alpha(\kappa,\tau) = \left[\begin{array}{l}
(g^+_\alpha+u^+_\alpha)\exp(+\frac{i}{2}\alpha) \\
(g^-_\alpha+u^-_\alpha)\exp(-\frac{i}{2}\alpha)  \end{array}\right]_{\TG},
\label{e:genspr}
\end{equation}
where the functions $g^\pm_\alpha(\kappa,\tau)$ and $u^\pm_\alpha(\kappa,\tau)$ are even and odd,
respectively, upon the reflection $\tau\to-\tau$. Here the superscripts $\pm$ are used for the
spinor components along $\TG$ to distinguish from the general notation $\uu\dd$. On a surface
of $C_{3v}$ symmetry, $g^+_\alpha$ and $u^-_\alpha$ are $2\pi/3$-periodic even functions of the
azimuthal angle $\alpha$, and $g^-_\alpha$ and $u^+_\alpha$ are odd functions of $\alpha$. In
order to construct a minimal model to approximately parametrize the variety of the data for
the different spin directions let us restrict ourselves to the first non-vanishing terms in
the Fourier expansion of the azimuthal dependence of the functions $g^\pm$ and $u^\pm$, i.e.,
a constant for the even and $\sin 3\alpha$ for the odd functions of $\alpha$:

\begin{equation}
\psi_\alpha(\kappa,\tau) = \left[\begin{array}{l}
(g^++u^+\sin 3\alpha)\exp(+\frac{i}{2}\alpha) \\
(g^-\sin 3\alpha+u^-)\exp(-\frac{i}{2}\alpha)  \end{array}\right]_{\TG}.
\label{modet}
\end{equation}

For example, in a circularly symmetric Rashba system the spin-orbit-split constant energy
contours form two concentric circles, one of which is described by Eq.~(\ref{modet}) with
$g^+$ being the only non-zero term, and the other one has the same structure of the spinor
for the opposite spin direction. On a $C_{3v}$ surface, for $\KK$ along the mirror plane
$\bar\Gamma\bar M$, the spinor components are of pure parity, i.e., in the general
equation~(\ref{e:genspr}) we have $u^+_0=0$ and $g^-_0=0$. This means that for the light
incident in the $xz$ plane a spin flip occurs on the change from $p$ to $s$ linear
polarization~\cite{Henk2003}.

Assuming the final state $\Phi_\alpha(\kappa,\tau)$ to be an $\alpha$-independent scalar
function even on reflection $\tau\to-\tau$ we arrive at the
angular dependence of the dichroism of the form $D_\tau(\alpha)\sim\sin 3\alpha$. It should
be however noted that the function $\Phi$ cannot be approximated by a single plane wave
$\exp(-i\KK\RR)$ because in this case the circular dichroism at normal incidence vanishes.
For the spin directions $\nn$ and $\KG$ this model yields the dichroism maps of a pure
$C_3$ symmetry, in agreement with the {\it ab initio} calculations in Figs.~\ref{f:ni1}
and~\ref{f:ni2}.

\begin{figure*}[ht] 
\centering
\includegraphics[width=0.95\textwidth]{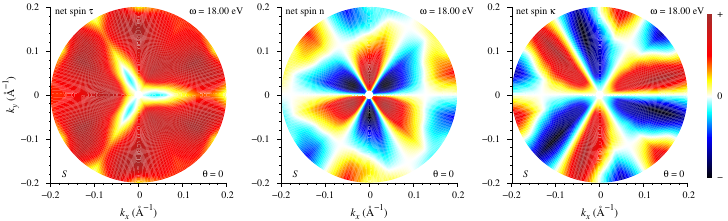} 
\caption{Spin polarization of the photocurrent at normal light incidence with a
  ``symmetrized'' polarization:
$S=I^{\uu}_{\rm CR}+I^{\uu}_{\rm CL}-I^{\dd}_{\rm CR}-I^{\dd}_{\rm CL}$.
  Left to right: $\sx\parallel\TG$, $\sx\parallel\nn$, and $\sx\parallel\KG$.
  Maps are normalized to the maximum magnitude.
}
\label{f:net}
\end{figure*}
Note that the simplistic model also has the property $D_\tau(\alpha)=-D_\tau(\alpha+\pi)$,
which is missing in the real surface. The violation of this relation is caused by the fact that
the final states $\Phi$ at $\KK$ and $-\KK$ are not related by symmetry, nor are they related
by time reversal because $\Phi$ and $\Phi^*$ propagate in opposite directions perpendicular to
the surface. In other words, the additional symmetry is brought about by ignoring the
$z$-component of the wave functions. In this regard it should be noted that the suggestion
in Ref.~\cite{ParkCH2012} to use for the final states the same time-reversal relation as for
the initial states is incompatible with the one-step-theory understanding of the final state
$\ket{{\rm f}\,\KK}$ as time-reversed LEED state~\cite{Feibelman1974}. (We note also that
a nearly-free-electron approximation for the final state, which assumes $\Phi$ to be a single
plane wave, does not change anything in this respect.) The ``apparent time-reversal symmetry
breaking'' was thoroughly experimentally studied in Ref.~\cite{Zhu2013} for \bss\ and analyzed
allowing for the three-dimensional nature of the surface states, and it was concluded to be
caused by a ``layer-dependent spin-orbital entanglement''. In fact, this effect is quite
general and occurs also without SOC.

This aspect has important implications also on the angular distribution of the spin
photocurrent. For example, in the measurement of Bi thin film on Si(111)~\cite{Takayama2011}
the in-plane spin-polarization of the photocurrent was found to be ``remarkably suppressed
on a half of six elongated hole pockets'' in apparent contradiction to the behavior of the
polarization of the initial states, see \eq{e:pol}, whose spinor wave functions for $\KK$
and $-\KK$ are related by time-reversal symmetry, so $S_{\TG}(-\KK)=S_{\TG}(\KK)$. As is clear
from the above discussion, the result of Ref.~\cite{Takayama2011} is in full accord with the
one-step theory, which we illustrate in the spin-polarization maps in Fig.~\ref{f:net}. In
order to retain the crystal symmetry in the numerical experiment, let us calculate the sum
of the net-spin photocurrents for CR and CL polarizations for a normal light incidence. The
$\sx\parallel\TG$ map is clearly seen to have the $C_{3v}$ symmetry, and the other two maps
are odd on reflection in the mirror planes. Interestingly, at $\hbar\omega=18$~eV for
$\sx\parallel\TG$ the spin-minority photocurrent dominates just at the mirror planes, but the
picture rapidly changes with photon energy: photocurrent from the same state may be spin-up
or spin-down depending on $\hbar\omega$. Another observation in Ref.~\cite{Takayama2011}
was that the out-of-plane net spin photocurrent was comparable to the in-plane one. The present
calculations are compatible with this result, but the actual intensity ratio depends on $\KK$
and on $\hbar\omega$: For example, for $\KA=0.14$~\AA$^{-1}$ the net spin amplitudes along
$\TG$ and $\nn$ are approximately equal at $\hbar\omega=8$--10~eV, and around 20~eV the
out-of-plane photocurrent becomes 10 times weaker, while the in-plane spin amplitude remains
of the same order of magnitude. Thus, our analysis fully supports the interpretation in
of the spin asymmetry Ref.~\cite{Miyamoto2018} as due to final-states effects. 
\begin{figure*}[ht] 
\centering
\includegraphics[width=0.95\textwidth]{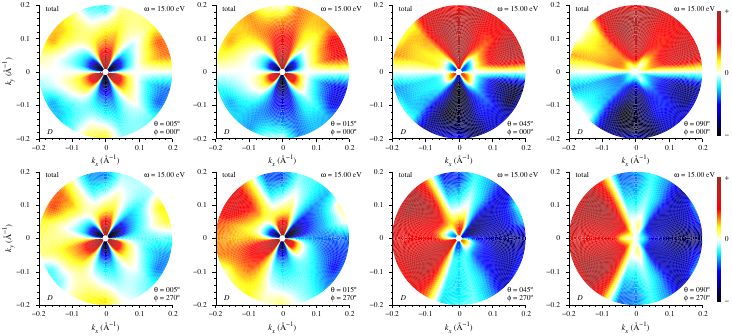} 
\caption{Off-normal incidence total-current CD for $\hbar\omega=15$~eV for two locations of the
  light incidence plane: $xz$ plane (upper row) and $yz$ plane (lower row). Four
  light-incidence angles are shown, left to right: $\theta=5^\circ$, 15$^\circ$, 45$^\circ$, and
  90$^\circ$. Maps are normalized to the maximum magnitude.
}
\label{f:off}
\end{figure*}

\subsection{Off-normal light incidence}

In a real experimental setup, the light cannot be directed perpendicular to the surface, so the
photoemission intensity depends on the location of the light incidence plane, angle $\phi$ in
Fig.~\ref{f:light}(a), and on the angle of incidence $\theta$, Fig.~\ref{f:light}(b). For
$\theta\ne0$, vector $\EG$ acquires a surface-perpendicular component, and the symmetry
of the angular distribution of the photocurrent becomes lower than in the $\theta=0$ case. 
Figure~\ref{f:off} shows the spin-unresolved CD for $\hbar\omega=15$~eV for the light incidence
along $\bar\Gamma\bar M$ and $\bar\Gamma\bar K$. In the former case, the experimental setup
retains the mirror symmetry across the $xz$ plane: $D(\alpha)=-D(-\alpha)$. Note, however, that
with increasing $\theta$ the angular distribution of CD gradually becomes positive everywhere
in the upper half-plane and negative in the lower one, i.e., $D(\alpha)$ approaches the
symmetry of $\sin\alpha$.

For the light incidence along $\bar\Gamma\bar K$ the CD maps possess no symmetry, but again,
with increasing $\theta$ its shape becomes more and more close to the symmetry of
$\sin(\alpha-\pi/2)$, i.e., positive almost everywhere in the left half-plane and
negative in the right one. Note that for an isotropic surface it holds
$D(\alpha)\sim\sin(\alpha+\phi)$, so the deviation from this pattern characterizes
the anisotropy of the electronic states involved. For example, the anisotropy of the surface
states on Cu(111) and Au(111) is much lower, and $D(\alpha)$ is rather close to the
$\sin\alpha$ shape~\cite{Kim2012OAM}. At the same time, tangible anisotropy is observed in
the $D(\alpha)$ of the Fermi contour of \bss~\cite{Park2012OAM}. Figure~\ref{f:off} shows that
for both $\phi$ angles, for $\theta=\pi/4$ the $D(\alpha)$ maps are almost the same as for the
grazing incidence, $\theta=\pi/2$. This means that the $\nn$-component of the momentum matrix
element is much larger than the in-plane components, and that the $\theta$-dependence of the
angular distribution of the azimuthal distribution of the photocurrent may give information
about their relative magnitude.

Regarding the spin dependence of CD at $\theta\ne0$, it follows from the spin structure
analysis in section~\ref{s:ground} that for the tangential spin $\sx\parallel\TG$, the
spin-majority CD$^+$ follows rather closely the $\sin(\alpha+\phi)$ pattern because the
majority spinor component is dominated by the even function $g^+_\alpha$, see~\eq{e:genspr},
whereas the spin-minority CD$^-$ has a more complicated angular dependence. 

\section{Conclusion}
To summarize, the presented calculations demonstrate that circular dichroism in photoemission
cannot be considered an intrinsic characteristic of the initial state even if the experimental
setup does not break the symmetry of the surface. A conspicuous manifestation of the problem is
the lack of a time-reversal relation between photoemission from $\KK$ and $-\KK$ in contrast to
its existence for the initial states. The present results indicate that the asymmetry between
$\KK$ and $-\KK$ observed in many experiments is not necessarily an artifact of the experimental
setup but is in full accord with the rigorous and well-grounded {\it ab initio} one-step theory.
The one-step formalism~\cite{Adawi64,Mahan1970,Caroli73,Feibelman1974,Pendry1976} makes quite
transparent the origin of the asymmetry: the presence of the time-reversed LEED state in
the matrix element, for which the relation $\Phi(\KK)=\Phi^*(-\KK)$ does not hold.

Similarly, the strong photon energy dependence of the CD maps for all the spin directions
considered and its
symmetry being lower than the symmetry of the initial state spin texture points to the absence
of a direct connection between CD and the spin polarization (which may appear as an artifact
of a simplified theoretical model~\cite{Wang2011}). This agrees with the conclusion of
Ref.~\cite{Scholz2013} regarding a minor role of spin in CD.

\section*{Acknowledgements}
This work was supported by the Spanish Ministry of Science and Innovation
(Projects No.~PID2022-139230NB-I00 and No.~PID2022-138750NB-C22).

%
\end{document}